\documentclass{article}
\usepackage{arxiv}
\usepackage[utf8]{inputenc}
\usepackage{hyperref}
 \usepackage{color}
 \usepackage{bm}
 \usepackage{graphicx, subcaption}
 \usepackage{latexsym}
 \usepackage{pifont}
 \usepackage{amssymb}
 \usepackage{hyperref}
 \usepackage{titlesec}
 \expandafter\let\csname equation*\endcsname\relax
 \expandafter\let\csname endequation*\endcsname\relax
 \usepackage{amsmath}
 \usepackage{listings}
 \usepackage{verbatim}
 \usepackage{geometry}
 \usepackage{booktabs}
 \usepackage{enumitem}
 \usepackage[normalem]{ulem} 
 \usepackage{breqn}
 \usepackage[toc,page]{appendix}
 \usepackage{cite}
 
 \definecolor{red}{rgb}{1.0,0.0,0.0}
 \definecolor{gre}{rgb}{0.0,1.0,0.0}
 \definecolor{blu}{rgb}{0.0,0.0,1.0}

 \definecolor{ora}{rgb}{1.0,0.5,0.0}
 \definecolor{gra}{rgb}{0.0,0.5,1.0}

 \newcommand{\change}[2]{#2}
 \newcommand{\changenew}[2]{#2}

 \newcommand{\ba}{\begin{abstract}}
 \newcommand{\ea}{\end{abstract}}
 \newcommand{\be}{\begin{equation}}
 \newcommand{\ee}{\end{equation}}

 \newcommand{\eg}{\textit{e}.\textit{g}. }

 \newcommand{\rom}[1]{\uppercase\expandafter{\romannumeral #1 \relax}}
 \newcommand{\Poincare}{Poincar$\acute{\rm e}$ }
 
 \newcommand{\Fig}[1]{figure \ref{fig:#1}}
 
 \newcommand{\Eqn}[1]{equation (\ref{eq:#1})}
 
 \renewcommand{\t}{\theta}
 \newcommand{\z}{\zeta} 
 \newcommand{\ds}{\displaystyle}
 \newcommand{\vect}[1]{\mathbf{#1}}
 \newcommand{\dd}[1]{\textnormal{d} #1}
 \newcommand{\pdv}[2]{\frac{\partial #1}{\partial #2}}

 \newcommand{\qty}{}
 \newcommand{\abs}[1]{|#1|}
 \newcommand{\grad}{\nabla}
 \newcommand{\tento}[1]{\times 10^{#1}}

\usepackage{physics} 
\usepackage{esint}
\usepackage{caption}
\usepackage[]{algorithm}
\usepackage[]{algorithmic}
\usepackage{xcolor}
\makeatletter
\def\mathcolor#1#{\@mathcolor{#1}}
\def\@mathcolor#1#2#3{%
  \protect\leavevmode
  \begingroup
    \color#1{#2}#3%
  \endgroup
}
\makeatother

\begin{document}
\title{Designing stellarators using perpendicular permanent magnets}
\author[*]{Caoxiang Zhu}
\author[ ]{Michael Zarnstorff}
\author[ ]{David Gates}
\author[ ]{Arthur Brooks}

\affil[ ]{Princeton Plasma Physics Laboratory, Princeton University, P.O. Box 451, New Jersey 08543, USA}
\affil[*]{\textit{Email: czhu@pppl.gov}}
\renewcommand\Authands{ and }
\date{}

\maketitle
\begin{abstract}
    We have developed a fast method to design perpendicular permanent magnets for simplifying stellarator coils based on existing codes. Coil complexity is one of the main challenges for stellarators. To date, only electromagnetic coils have been used to generate 3D fields for stellarators. Permanent magnets provide an alternative way to produce the desired magnetic field for optimized stellarators. In this paper, we revisit the concept of representing surface current using magnetic dipoles and carry out numerical validations. A surface magnetization is proven to be equivalent to the surface current that can be linearly solved by existing coil design codes. An incremental multi-layer method has been developed to obtain a practical solution that is attainable with present permanent magnets. With this method, we can reproduce a half-Tesla NCSX configuration using specially designed neodymium magnets together with simple planar coils. It shows that stellarator coils could be substantially simplified by adopting permanent magnets.
\end{abstract}

\section{Introduction}
A stellarator is a toroidal magnetic confinement configuration that uses external coils to produce a non-axisymmetric magnetic field for confining the plasma.
It is an attractive approach to fusion energy because stellarators have low recirculating power and are free of disruptions.
On the other side, the 3D nature of stellarators generally requires more complicated coils than axisymmetric configurations.
Complex coils are one of the main challenges for stellarators.
The dominant cost growth factors of the National Compact Stellarator eXperiment (NCSX) project were found to be complicated geometries and the associated accuracy requirements \cite{Strykowsky2009}.
Difficulties in fabricating and assembling three-dimensional non-planar coils were also recognized during the construction of Wendelstein 7-X (W7-X) experiment \cite{Bosch2013c}.

In recent years, substantial efforts have been devoted to simplifying stellarator coils.
Novel coil design methods, like the straight out-leg coil designs \cite{Gates2017}, REGCOIL\cite{REGCOIL}, coil winding surface optimization method \cite{Paul2018} and 3D nonlinear optimization code FOCUS \cite{FOCUS00} are developed.
New numerical techniques using shape gradient \cite{Landreman2018}, stochastic optimization \cite{Lobsien2018} and Hessian matrix method \cite{Hessian01} are also proposed to better identify or relax coil tolerance.
In addition, integrated optimization on both plasma and coils is explored to find configurations with simpler coils \cite{Hudson2018}.
These studies are all concentrating on current-carrying electromagnet (normal or superconducting), as it is the only type that has been used on stellarators to date.

Recently, the use of permanent magnets has been proposed to produce the three-dimensional component of stellarator magnetic fields and simplify the required electromagnetic coils \cite{HelanderPM}.
Besides electromagnet, permanent magnet is the most ancient, and perhaps the most common, way to generate a magnetic field.
Permanent magnets are used in daily life, industries and laboratories.
Advances in material science have brought new magnetic materials that have relatively strong residual flux density (Br), \eg the Nd-Fe-B magnet has been reported to have a record of Br=1.555 T \cite{NdFeBRecord}.
Although this is rather modest compared to electromagnets especially considering that the strongest field outside the magnet is normally half of the residual field, special arrangements of permanent magnets, which are normally called Halbach arrays \cite{Halbach1980}, can generate a much higher field.
For instance, a 5.16T magnetic field was measured in a specialized permanent dipole magnet using Nd-Fe-B materials \cite{Kumada2004}.
Such a field is higher than many existing magnetically confined fusion experiments.
The reason why permanent magnet might be of great potential for stellarators is that it could provide a steady magnetic field without energy consumption and more importantly, the material is considerably inexpensive, which possibly leads to a significant cost reduction. 
\change{}{On the other hand, this could also be a disadvantage since the magnets cannot be turned off.}

\change{}{Designing permanent magnets for stellarators can be interpreted as the inverse problem for the Biot-Savart law, namely how to determine the magnetization for a given magnetic field.
It is the intermediate step between physics design and engineering design.
Therefore, a desired configuration should be provided as the target and only limited engineering constraints will be considered during this step.
For instance, the magnetization used should be attainable by present material.
In addition, permanent magnets are sensitive to temperature and \changenew{might}{will} be demagnetized by neutrons \cite{Alderman2002}, so they have to be placed outside the vacuum vessel (as well as the blanket if needed for neutron shielding).
}

\change{The}{} Ampere's law states that $\ds \oint_C \vect{B} \cdot \dd{\vect{l}} = \mu_0 I_{free} + \oint_C \mu_0 \vect{M} \cdot\dd{\vect{l}}$, where $\vect{B}$ is the total magnetic field (magnetic flux density), $\vect{M}$ the magnetization, $C$ an arbitrary closed curve, $I_{free}$ the electric current enclosed by the curve.
If $C$ is chosen to be \changenew{a toroidal closed curve lying on the inboard side of the plasma boundary}{the magnetic axis inside the plasma boundary}, the magnetization part in RHS should be zero while the LHS is usually non-zero in toroidal devices. 
This means that the free current through the hole of \change{}{the} torus (poloidal current), $I_{free}$, is non-zero.
In other words, electromagnets (coils) are always required.
This is not a problem as we are not anticipating to using permanent
magnets providing the entire field.
Hopefully, stellarator coils can be simplified by employing permanent magnets to provide the complementary 3D field.
This idea was briefly mentioned by Ku \& Boozer \cite{Ku2010}, where they proposed using dipoles to contribute part of the magnetic field but wasn't explored further.
Later, Boozer \cite{Boozer2015Review} pointed out that the single-valued part of the current potential can be interpreted as the number of dipole moments per unit area.
Helander \etal \cite{HelanderPM} introduced an approach that uses a curl-free `one-sided' magnetization which is tangent to a toroidal surface.
Together with planar coils, the specially calculated magnetization can produce the required magnetic field for optimized stellarators. 
In this paper, we are going to numerically verify that \change{the}{a} surface magnetization from \change{}{the} current potential can generate the desired magnetic field. 
Furthermore, we will introduce a multi-layer method to design optimized stellarators using perpendicular permanent magnets with finite thickness and the maximum magnetization is attainable with present material technology. 

The paper is organized as follows. 
In Sec. \ref{method}, we briefly review the least-square minimization
problem of coil designs and interpret how this would lead to a
solution to surface magnetization.
A multi-layer direct construction method to find solutions that are practical to
real magnet materials is also introduced.
In Sec. \ref{ellipse}, numerical validations are carried out on a
conventional rotating elliptical configuration to check the
convergence properties.
We introduced a novel design with only TF coils together with
perpendicular permanent magnets that produces the desired magnetic field for a half-Tesla NCSX equilibrium in \ref{NCSX}.
We will summarize in Sec. \ref{discussion}.

\section{Equivalent magnetization to surface current} \label{method}
\subsection{Revisit of the surface current method for designing stellarator coils}
Given the current distribution, it is easy to compute the generated magnetic field by following the Biot-Savart law.
The inverse problem is much more complicated.
Pioneering work was done by Merkel \cite{NESCOIL} using Green's functions with the development of NESCOIL code.
On a prescribed outer surface surrounding the plasma, a divergence-free surface current density $\vect{K}$ \change{}{(unit: A/m)} is represented by,
\begin{equation} \label{eq:current_density}
    \vect{K} = \vect{n} \times \grad{\Phi} \ ,
\end{equation}
where $\Phi$ is the surface current potential \change{}{(unit: A)} and $\vect{n}$ the unit normal vector of the current-carrying surface (winding surface).
The surface current distribution is calculated by minimizing the normal magnetic field on the plasma surface (Neumann boundary condition), 
\begin{equation} \label{eq:chi2B}
    \chi^2_B = \iint_{\partial P} (\vect{B} \cdot \vect{n'})^2 \dd{S'} \ ,
\end{equation}
\change{}{where $\vect{n'}$ is the unit normal vector of the plasma surface.
Hereafter, primes are used to denote coordinates on the plasma surface and the winding surface un-primed.}
In \Eqn{chi2B}, the total magnetic field consists of different
components, $\vect{B} = \vect{B}_{plasma} + \vect{B}_{fixed} +
\vect{B}_{K}$, where $\vect{B}_{plasma}$ is arising from plasma
currents provided by MHD equilibrium codes, $\vect{B}_{fixed}$ from
fixed external coils (like toroidal field coils) and $\vect{B}_{K}$
from the surface current to be solved. $\vect{B}_{K}$ is calculated
using the Biot-Savart law,
\begin{equation}
\vect{B}_{K} = \frac{\mu_0}{4\pi} \iint_{\partial C} \frac{\vect{K} \times (\vect{x}' - \vect{x}) }{\abs{\vect{x}' - \vect{x}}^3} \dd{S} \ .
\end{equation}
Using Green's function the determination of the surface current (current potential) becomes a least-square minimization problem and can be solved linearly \cite{NESCOIL, REGCOIL}.
Afterwards, discrete coils are cut by following the contours of $\Phi$.

The inverse problem is ill-posed.
NESCOIL truncates a finite number of Fourier modes in representing $\Phi$ and NESVD \cite{Pomphrey2001} uses singular-valued decomposition (SVD) while REGCOIL adopts a Tikhonov regularization over the surface integral of the squared current density.

\subsection{Equivalent surface magnetization to surface current}
The vector potential produced by a surface current $\vect{K}$ is calculated as
\begin{equation} \label{eq:bs_surface}
    \ds \vect{A}_K = \frac{\mu_0}{4\pi} \iint_{\partial C}  \frac{\vect{K}}{r} \dd{S} \ ,
\end{equation}
where $r=\abs{\vect{x}' - \vect{x}}$ is the position vector.
Substituting \Eqn{current_density} into \Eqn{bs_surface}, we have
\begin{align} \label{eq:AK}
    \ds \vect{A}_K = \frac{\mu_0}{4\pi} \iint_{\partial C} \frac{\Phi \vect{n} \times \vect{r}}{r^3} \dd{S} \ .
\end{align}
(Details of the derivation can be found in Appendix \ref{derivation}.)

\change{Suppose there is surface magnetization lying on the winding surface and pointing in normal direction}
{The strength of permanent magnets can be expressed by magnetization,  $\vect{M}$ (unit: A/m), which is the quantity of magnetic moment,  $\vect{m}$ (unit: $\mathrm{A \cdot m^2}$), per unit volume.
Suppose there is a magnetization with zero thickness, namely a ``surface magnetization''.
The surface magnetization lies on the winding surface and its orientation is along the surface normal.
We can express the surface magnetization with a Dirac delta function in the normal direction}, $\vect{M} = \Phi \vect{n}\delta(s-s_0)$, where $(s, \t, \z)$ forms a curvilinear coordinate system with $s$ in \change{}{the} normal direction, $\theta$ the poloidal angle and $\zeta$ the toroidal angle.
The winding surface locates at $s = s_0$.
This surface magnetization could be discretized into localized magnetic dipoles where the magnetic moment in each dipole is computed as
\begin{equation} \label{eq:dipole}
    \ds \vect{m}_{i_{\t},i_{\z}} = \iiint \vect{M} \dd{V} = \int_{s}  \delta(s-s_0) \dd{s} \int_{\t} \int_{\z} \Phi \vect{n} \sqrt{g} \dd{\t} \dd{\z} = [{\Phi \vect{n} \Delta{S}}] _{i_{\t},i_{\z}}\ ,
\end{equation}
\change{}{where $\Delta{S}$ is the area of the surface element.
Here, we assume the magnetization is homogeneous inside each discrete element \changenew{}{and write the quantities in discretized forms}.}
The vector potential produced by a single magnetic dipole is computed as,
\begin{equation}
    \vect{A}_{i_{\t},i_{\z}} = \frac{\mu_0}{4\pi} \frac{\vect{m}_{i_{\t},i_{\z}} \times \vect{r}}{r^3} \ ,
\end{equation}
\changenew{}{where $\vect{r}$ is the position vector from the dipole origin to the evaluation point and for clarity we omit the subscripts ${i_{\t}}$ and ${i_{\z}}$. }
Following the ansatz of surface magnetization and discrete magnetic dipoles, the total vector potential from the surface magnetization is
\begin{equation} \label{eq:AM}
    \ds \vect{A}_M = \sum_{i_{\t}} \sum_{i_{\z}} \vect{A}_{i_{\t},i_{\z}} = \frac{\mu_0}{4\pi} \sum_{i_{\t}} \sum_{i_{\z}} \frac{\Phi \vect{n} \times \vect{r}}{r^3} \Delta{S} \ .
\end{equation}
Equation (\ref{eq:AM}) is essentially the discretized form of \Eqn{AK}, which implies that a surface magnetization on the winding surface following the distribution of current potential and positioning normally will produce the same magnetic field as a surface current density up to the accuracy of discretization.

With the surface current potential calculated by NESCOIL (NESVD or REGCOIL), one can easily obtain a surface magnetization distribution that will produce the desired magnetic field required for plasma equilibrium.
This only requires solving linear equations, so it is fast and avoids trapping in local minima. 

\subsection{Permanent magnets with finite thickness} \label{multi}
The surface magnetization can be well approximated by a thin layer of permanent magnets.
But there is a technical limit on the maximum magnetization that can be produced in a specific magnetic material.
This is often expressed in the form of the residual flux density (remanent field) $B_r$.
The dipole moment is calculated by the volume integral of the
magnetization, $\vect{m} = (\vect{B}_r / \mu_0) V$.
Therefore, to provide \change{}{an} equivalent amount of magnetic moment, one should
use a permanent magnet with the  thickness (in \change{}{the} normal direction) of $h = \mu_0 \Phi / B_r$.
When using Nd-Fe-B magnets and taking $B_r=1.4$T as being for inexpensively available material, the required thickness is determined by the value of current potential with the formula $h=\Phi/(1.1\tento{6})$ meter.

In \change{}{the} above ansatz, we are using the surface magnetization result and ignoring the effect of thickness.
It is a good approximation only if the thickness of permanent magnets is much smaller \change{that}{than} the distance from the magnets to the plasma surface.
This condition is not always satisfied, as the current potential varies in different configurations with different choices of the winding surface.
However, it is still possible to obtain arrangements of permanent magnets with finite thickness by using the following method.

Instead of using one single winding surface, we choose multiple nested surfaces labeled as $S_1, S_2, \cdots, S_N$ from innermost to outermost.
For each surface, \change{}{an} additional constraint on the maximum current
potential is imposed to make sure the magnetization is within the material limit.
For instance, $\Phi_{m} = 1.1\tento{3}$A is the limit of  the surface
magnetization for 1-mm-thick Nd-Fe-B magnet and 1 mm would be sufficiently thin to be approximated by surface magnetization.
Starting from the innermost surface $S_1$, one can obtain a surface magnetization $\vect{M}_1$ that minimizes normal field error $\chi^2_{B}$ in \Eqn{chi2B} subjected to the constraint $\Phi_{max} \leq \Phi_{m}$ by varying the regularization term (the Fourier resolution in NESCOIL, the number of SVD truncated terms in NESVD, the squared current density in REGCOIL or the surface integral of squared current potential as in Appendix \ref{phi_regularize}).
One can also solve the current potential $\Phi_1$ normally without limiting $\Phi_{max}$ and then truncate the current potential to $\Phi_m$ when extracting the surface magnetization $\vect{M}_1$.
Because of the constraint/truncation, $\chi^2_{B}$ might not be sufficiently small.
The magnetic field produced by $\vect{M}_1$, denoted as $\vect{B}_1$, is then calculated and considered as part of the ``fixed'' magnetic field.
For the next surface $S_2$, optimal $\vect{M}_2$ is obtained with the same constraint/truncation and again the produced magnetic field $\vect{B}_2$ is added to the fixed magnetic field.
By incrementally adding more surfaces, the resulting $\chi^2_{B}$ will be lower and lower until it converges (or below some allowable value).
Therefore, we can obtain the distribution of magnetization for permanent magnets with finite thickness, more importantly, within the availability of present magnet materials.
The multi-layer implementation using truncation is described in algorithm \ref{alg1}.
\begin{algorithm}
\caption{Obtain permanent magnets with finite-thickness.}
\label{alg1}
\begin{algorithmic}
\STATE $i \leftarrow  1$
\STATE $\Phi_m \leftarrow B_r/\mu_0 \,h$, where $h \ll d_{plasma-magnets}$
\STATE $\vect{B}_{fixed} \leftarrow  \vect{B}_{coils}$
\WHILE{$\chi^2_{B}>$ target\_value}
\STATE prepare winding surface $S_i$
\STATE $\ds \min_{\Phi_i} \quad  \iint_{\partial P} \qty[ \qty(\vect{B}_{plasma} + \vect{B}_{fixed} + \vect{B}(\Phi_i)) \cdot \vect{n} ]^2 \dd{S} + \lambda \chi^2_{R}$
\IF{$\Phi_i(\t, \z)<\Phi_m$}
\STATE $\vect{M}_i (\t, \z) \leftarrow \Phi_i(\t, \z) \vect{n}_i \delta(s-s_i)$
\ELSE
\STATE $\vect{M}_i (\t, \z) \leftarrow \Phi_m \vect{n}_i \delta(s-s_i)$
\ENDIF
\STATE discretize and calculate $\vect{B}(\vect{M}_i)$
\STATE $\vect{B}_{fixed} \leftarrow  \vect{B}_{fixed} + \vect{B}(\vect{M}_i)$
\STATE $i \leftarrow  i+1$
\ENDWHILE
\end{algorithmic}
\end{algorithm}

\section{Numerical validations} \label{ellipse}
\begin{figure}
    \centering
    \includegraphics[width=0.8\textwidth]{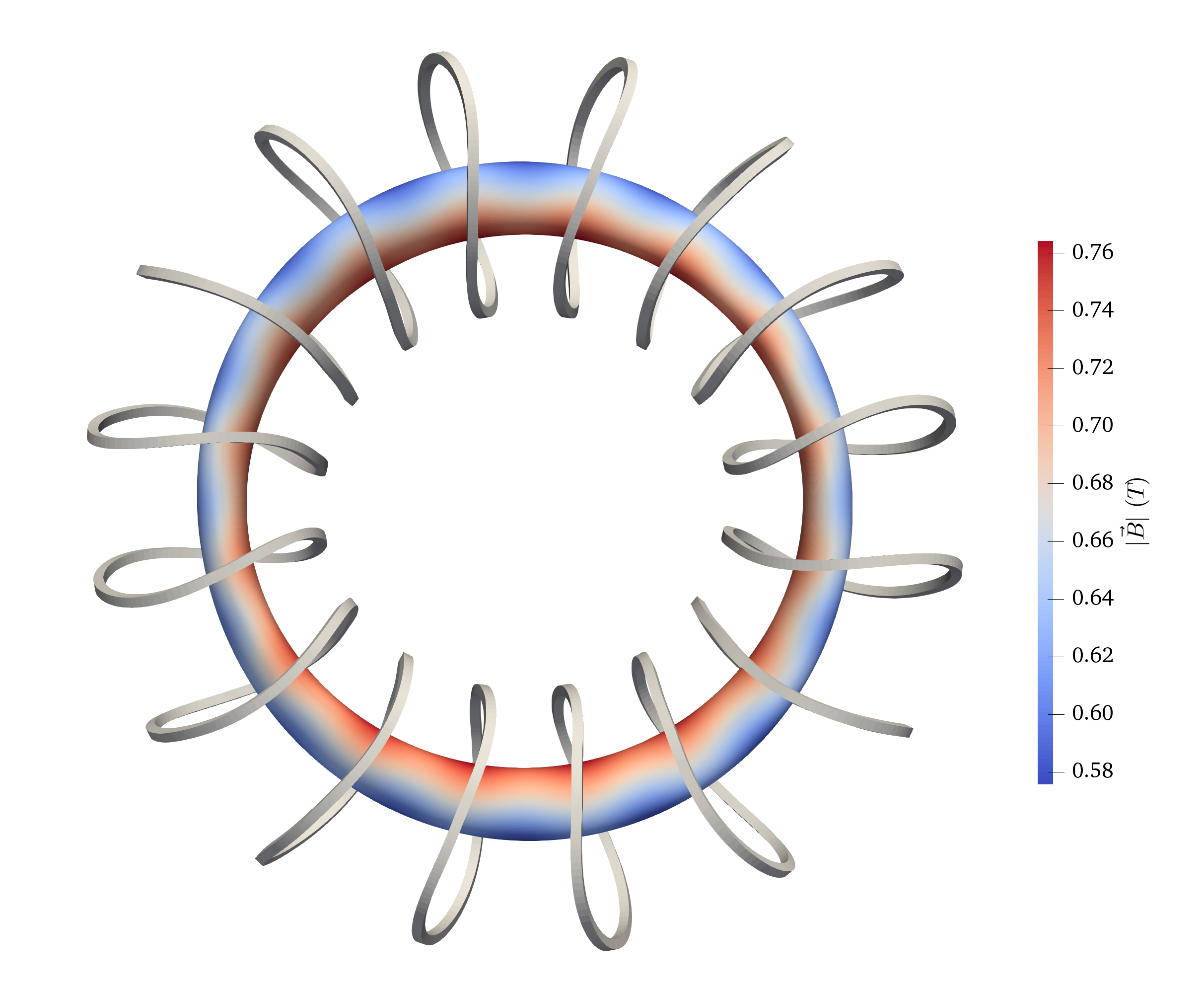}
    \caption{A classical $l=2$ stellarator and modular coils optimized by FOCUS. Colors on the plasma surface indicate the magnitude of magnetic field produced by modular coils.}
    \label{fig:ellipse_coils}
\end{figure}

To numerically validate the proposed method, we choose a classical $l=2$ toroidal stellarator.
The plasma boundary is a two-period rotating ellipse, described in cylindrical coordinates ($R, \z, Z$) by Fourier harmonics as
\begin{equation}
    \begin{cases}
    R = 3.0 + 0.3 \cos(\t) -0.06 \cos(\t-2\z) \ ;\\
    Z = 0.3 \sin(\t) + 0.06 \sin(-2\z) + 0.06 \sin(\t-2\z) \ .
    \end{cases}
\end{equation}
Here we shall just consider the vacuum case ($\vect{B}_{plasma}=0$).
Normally, this configuration can be built with helical coils or modular coils (non-planar coils that can be independently manufactured and removed).
In \Fig{ellipse_coils}, a modular coil design with 16 coils \change{are}{is} shown.
These coils are represented with five Fourier modes in each coordinate and optimized by the coil design code FOCUS \cite{FOCUS00} with a small weight on the coil length penalty.
Coil currents are fixed to be 0.625 MA such that the total poloidally linked current $I_{coils}$ is 10 MA.
The residual normal field error $\chi^2_{B}$ is $6.33\tento{-4} \mathrm{T^2m^2}$. 

\subsection{Surface magnetization solution}
\begin{figure}
\centering
\begin{minipage}[t]{.49\linewidth}
  \centering
  \includegraphics[width=\linewidth]{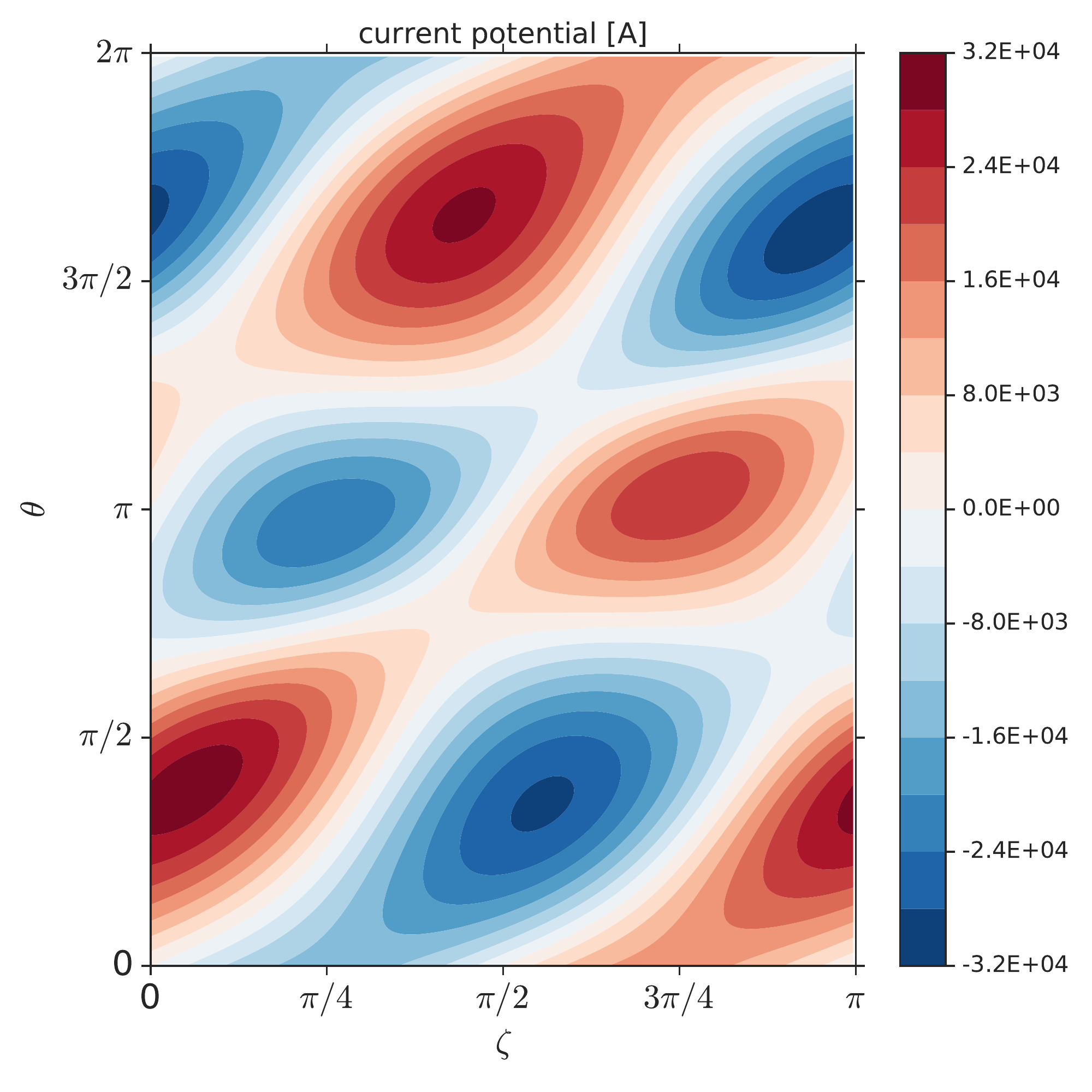}
  \caption{Contours of current potential solved on the winding surface in \Fig{winding_surface} with $\lambda_{K}=1.0\tento{-23}$. Only one toroidal period is shown here.}
  \label{fig:cp_contour}
\end{minipage}\hfill
\begin{minipage}[t]{.49\linewidth}
  \centering
  \includegraphics[width=\linewidth]{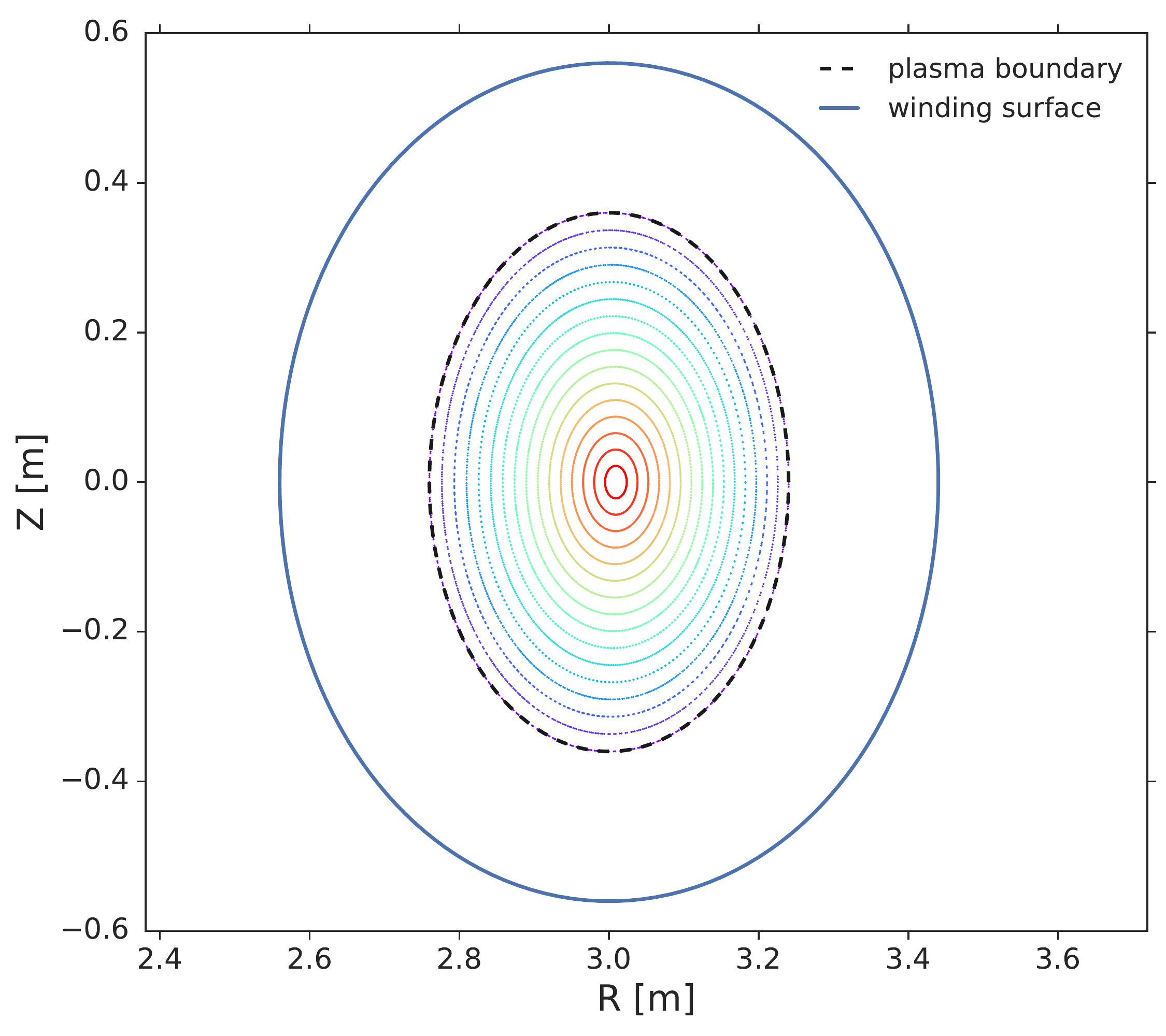}
  \caption{\Poincare plots at the $\z=0$ plane from field-line following in the magnetic field produced by the central current and $128\times128$ magnetic dipoles on the winding surface. Cross-sections of the plasma boundary (dash lines) and the winding surface (solid lines) are also shown.}
  \label{fig:winding_surface}
\end{minipage}
\end{figure}

As discussed above, permanent magnets cannot provide \change{}{a} poloidally linked current.
\change{A vertical current}{An infinitely long wire} ($I_{coils}=10$MA) at the center of torus was used to produce the toroidal field which is proportional to $1/R$.
In axisymmetric devices, like tokamaks, such a toroidal field is often generated by simple planar toroidal field (TF) coils.
The winding surface was generated by uniformly expanding the plasma surface by a distance of 0.2 m, as shown in \Fig{winding_surface}.
We used REGCOIL to calculate the current potential with Fourier resolution of $M_F=20, N_F=20$ and spatial resolution of $N_{\t} \times N_{\z} = 128 \times 128 $ (per period) for both the winding surface and the plasma surface.
The regularization parameter of the current density is set to $\lambda_{K}=1.0\tento{-23}$.
The maximum current potential is about $3.2\tento{4}$ A and the resulting $\chi^2_B$ is $2.01\tento{-16} \mathrm{T^2m^2}$.
Figure \ref{fig:cp_contour} shows the distribution of current potential on the ($\t,\z$) surface.
Afterwards, we extracted discrete magnetic dipoles following
\Eqn{dipole} with a resolution of $N_d^{\t} \times N_d^{\z} = 128 \times 128$ \change{}{along the structured grid.
The origin of each dipole locates at the intersection of the poloidal and toroidal coordinate curves.
The orientation is perpendicular to the surface and the area of each surface element is calculated as $\Delta{S}_{i_{\t},i_{\z}} = |\pdv{\vect{r}}{\t} \cross \pdv{\vect{r}}{\z}| \Delta{\t} \Delta{\z}$}.
We modified FOCUS to be capable of calculating the magnetic field from magnetic dipoles.
The residual normal field error from the central wire and discrete magnetic
dipoles is  $5.20\tento{-16} \mathrm{T^2m^2}$.
Field-line tracing results show that the vacuum flux surface is almost identical to the target magnetic surface, as illustrated in \Fig{winding_surface}. The relative difference of the rotational transform at the last flux surface is only $0.05\%$.

%

\begin{figure}
\centering
\begin{minipage}[t]{.49\linewidth}
    \centering
    \includegraphics[width=\linewidth]{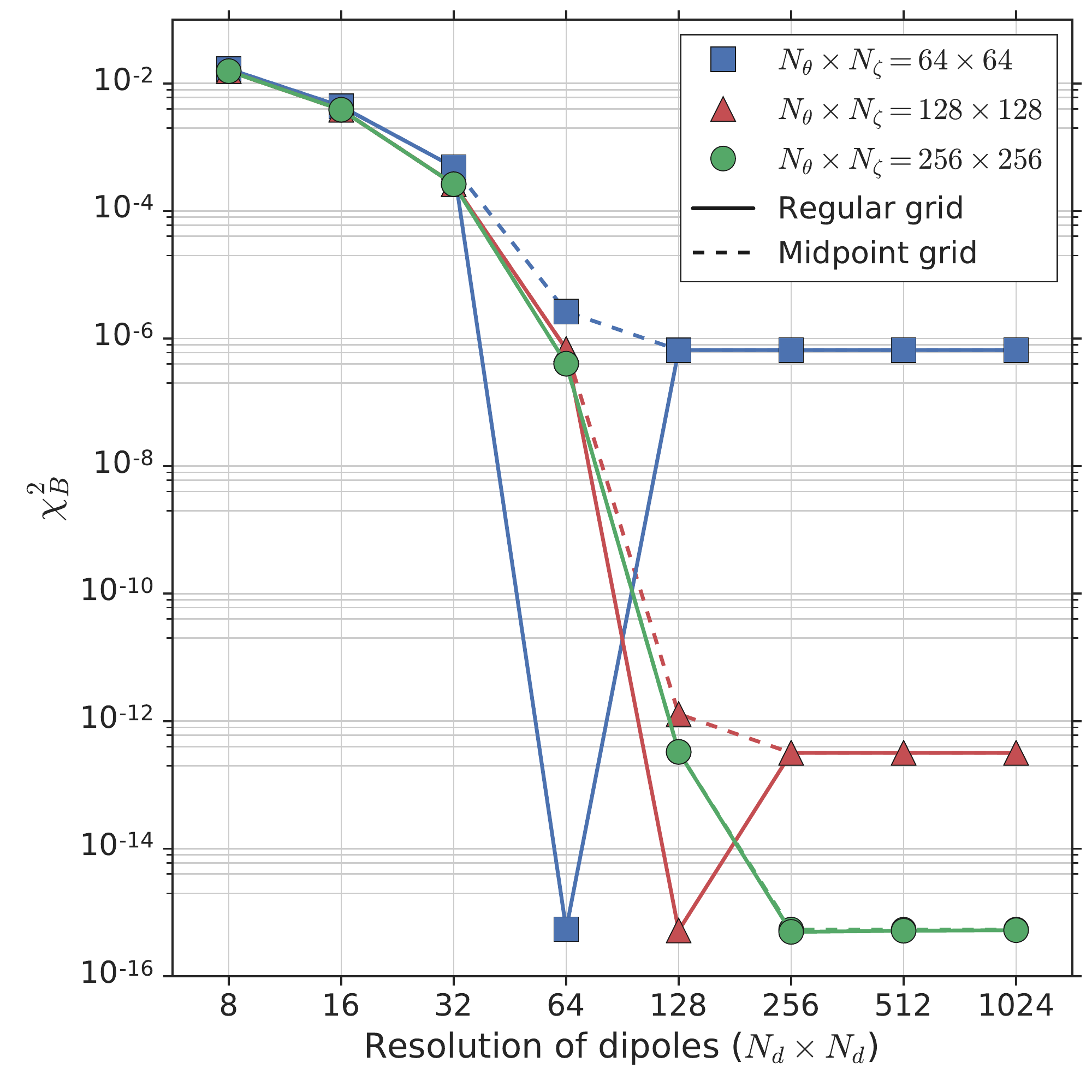}
    \caption{Convergence with different number of magnetic dipoles discretized from three current potentials. \change{}{using both the ``regular'' grids and ``midpoint'' grid. Magnetic dipoles are discretized at the intersection of $\t$ and $\z$ coordinate for the regular grid cases and at the midpoint for the midpoint grid cases.} $\chi^2_B$ is calculated following \Eqn{chi2B} while the total magnetic field is from the central current and discretized magnetic dipoles.}
    \label{fig:single_convergence}
\end{minipage}\hfill
\begin{minipage}[t]{.49\linewidth}
    \centering
    \includegraphics[width=\linewidth]{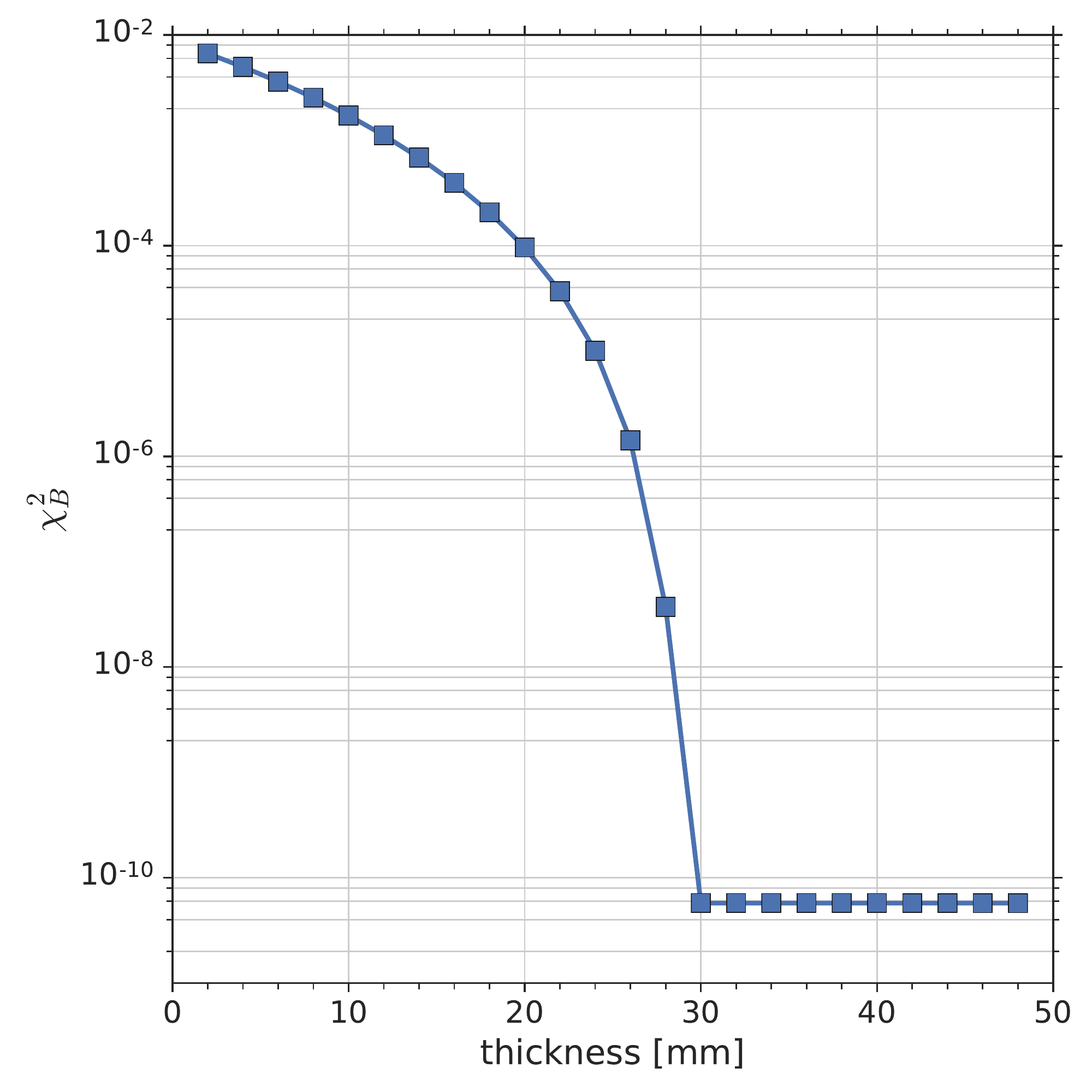}
    \caption{Convergence with multi-layer implementation.}
    \label{fig:multi_convergence}
\end{minipage}
\end{figure}

Convergence studies were also performed for magnetic dipoles on \change{}{a} single surface.
We obtained three current potential distributions calculated with spatial resolution $N_{\t} \times N_{\z} $ of $64 \times 64$, $128 \times 128$ and $256 \times 256$ using REGCOIL with the same regularization parameter $\lambda_{K}=1.0\tento{-23}$.
For each current potential, we discretized different numbers of magnetic dipoles\change{}{, as explained above}.
\change{}{We also tried to discretize the dipoles by using the midpoint of the grids.}
The residual normal field error $\chi^2_B$ was then evaluated using FOCUS in each case.
As shown in \Fig{single_convergence}, $\chi^2_B$ is generally smaller with more dipoles.
\change{If the resolution of dipoles is identical to the spatial resolution used for calculating the current potential, we will recover exactly the same precision as using the regularized surface currents from REGCOIL.
This minimum values for different current potentials are at the same order, which is consistent with REGCOIL results.
However, the minimum solutions are numerically unstable.
When the resolution of dipoles exceeds the resolution used for current potential, $\chi^2_B$ might become higher.}
{$\chi^2_B$ for the case of $N_{\t} \times N_{\z} = 64 \times 64$ converges at the order of $10^{-7}$, $N_{\t} \times N_{\z} = 128 \times 128$ at $10^{-13}$ and $N_{\t} \times N_{\z} = 256 \times 256$ at $10^{-16}$.
There are two abnormal data points.
When the dipoles are discretized at the intersection of the $\theta$, $\zeta$ coordinate (``regular grid''), as the same discretization in REGCOIL, we will recover exactly the same precision for $\chi^2_B$ ($\sim 10^{-16}$)  as REGCOIL if the resolution of dipoles is identical to the spatial resolution used for calculating the current potential.
It is not observed when the dipoles are discretized at the midpoint of the grids (``midpoint grid''), even if $N_d^{\t} \times N_d^{\z} = N_{\t} \times N_{\z}$.}
It indicates that the calculated current potential has a dependence on the discretization.
The calculated Fourier coefficients for current potentials are also different in the three cases.
The explanation is that we are using a least-squares minimization method to calculate the Fourier coefficients for current potentials.
The higher resolution used, the \change{better}{more robust} the solution is.
If we use sufficiently high resolutions for calculating the current potential, like $N_{\t} \times N_{\z} = 256 \times 256$ in this case, we can obtain a stable solution.

%

\subsection{Magnetization with finite thickness}
\begin{figure}
    \centering
    \includegraphics[width=0.8\textwidth]{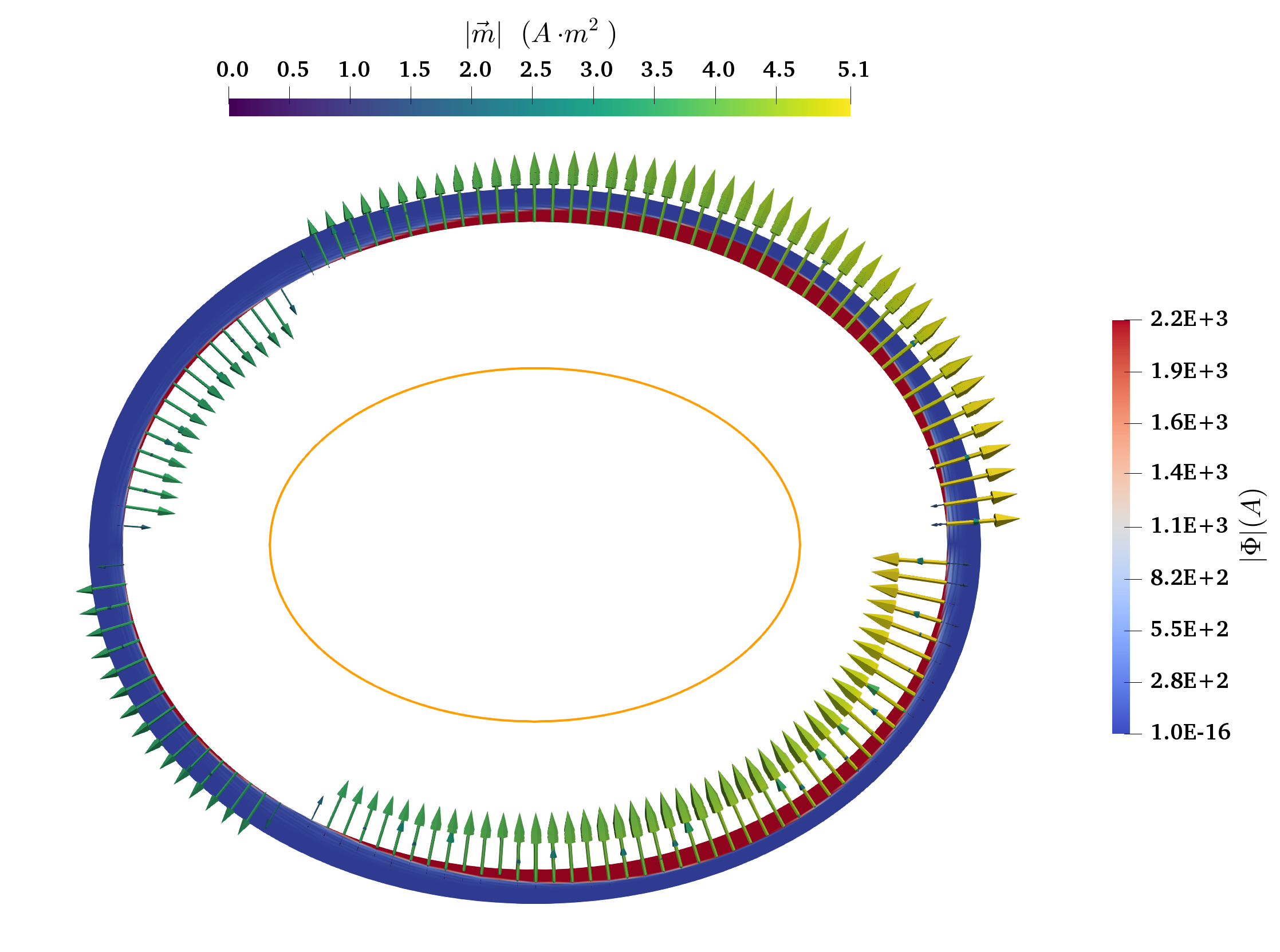}
    \caption{Distribution of current potentials and the magnetic moments at $\z=\pi/2$ plane. The arrows in different radial layers are \change{stacking}{staked} together.}
    \label{fig:ellipse_multi}
\end{figure}

In the above calculation, the maximum current potential is $3.2\tento{4}$ A, which is equivalent to \change{a surface magnetic moment}{the magnetic moment per unit area} of 2.91-cm-thick Nd-Fe-B permanent magnet.
The distance from the winding surface to the plasma boundary is 20cm.
Hence, the surface magnetic moment approximation would be adequate. 
Of course, we can still demonstrate the multi-layer implementation proposed in Section \ref{multi}.
Here, we chose a sequence of nested surfaces, starting from the same winding surface we used for the single-layer calculation ($\Delta r = 20$  cm).
The surfaces are all obtained by uniformly expanding the plasma
boundary while the distance between two adjacent surfaces is 2 mm.
Therefore, the maximum allowable current potential on each layer, $\Phi_m$ is about $2.2\tento{3}$ A.
Following algorithm \ref{alg1}, we can incrementally get the distribution of magnetization.
\changenew{}{We used the resolution of $N_d^{\t} \times N_d^{\z} = N_{\t} \times N_{\z} = 128 \times 128$ and discretized magnetic dipoles on the regular grids for each layer.}
In \Fig{multi_convergence}, the residual Bn error is reducing as the
thickness increases and it converges when the total thickness is 3.0
cm, which is consistent with the surface magnetization approximation.
The overall distribution of the multi-layer magnetization at one cross-section is shown in \Fig{ellipse_multi}.

\section{Magnetization for the half-Tesla NCSX configuration}\label{NCSX}
The rotating elliptical stellarator is a simple model for validation.
Here we can also apply this method to a practical optimized stellarator, NCSX \cite{Zarnstorff2001}.
NCSX was optimized to have good quasi-axisymmetry and be MHD stable.  
Modular coils \cite{Williamson2005} were designed to provide the main
magnetic field, while TF and PF coils are used to increase flexibility.
We chose one of the designed NCSX equilibria, C09R00, as our reference equilibrium.
Some key parameters of C09R00 are $\mathrm{N_{fp}} = 3$, $R_0=1.44$ m,
$a=0.32$ m, $V_{\mathrm{plasma}}=2.96 \mathrm{m^3}$, $\langle B
\rangle=1.57$ T and $\langle \beta \rangle=4.09\%$.
To demonstrate that permanent magnets can simplify coil designs, we will only keep the planar TF coils to provide the essential toroidal field.
As the existing TF coils were not built to produce a magnetic field as high as 1.57T, we scaled the volume\change{}{-}averaged magnetic field to 0.5T that the TF coils can sustain.
The pressure profile and plasma currents were also scaled down to retain the same $\langle \beta \rangle$, while the other parameters, like the shape and the iota profile, were not changed.

\begin{figure}
\centering
\begin{minipage}[t]{.49\linewidth}
    \centering
    \includegraphics[width=\linewidth]{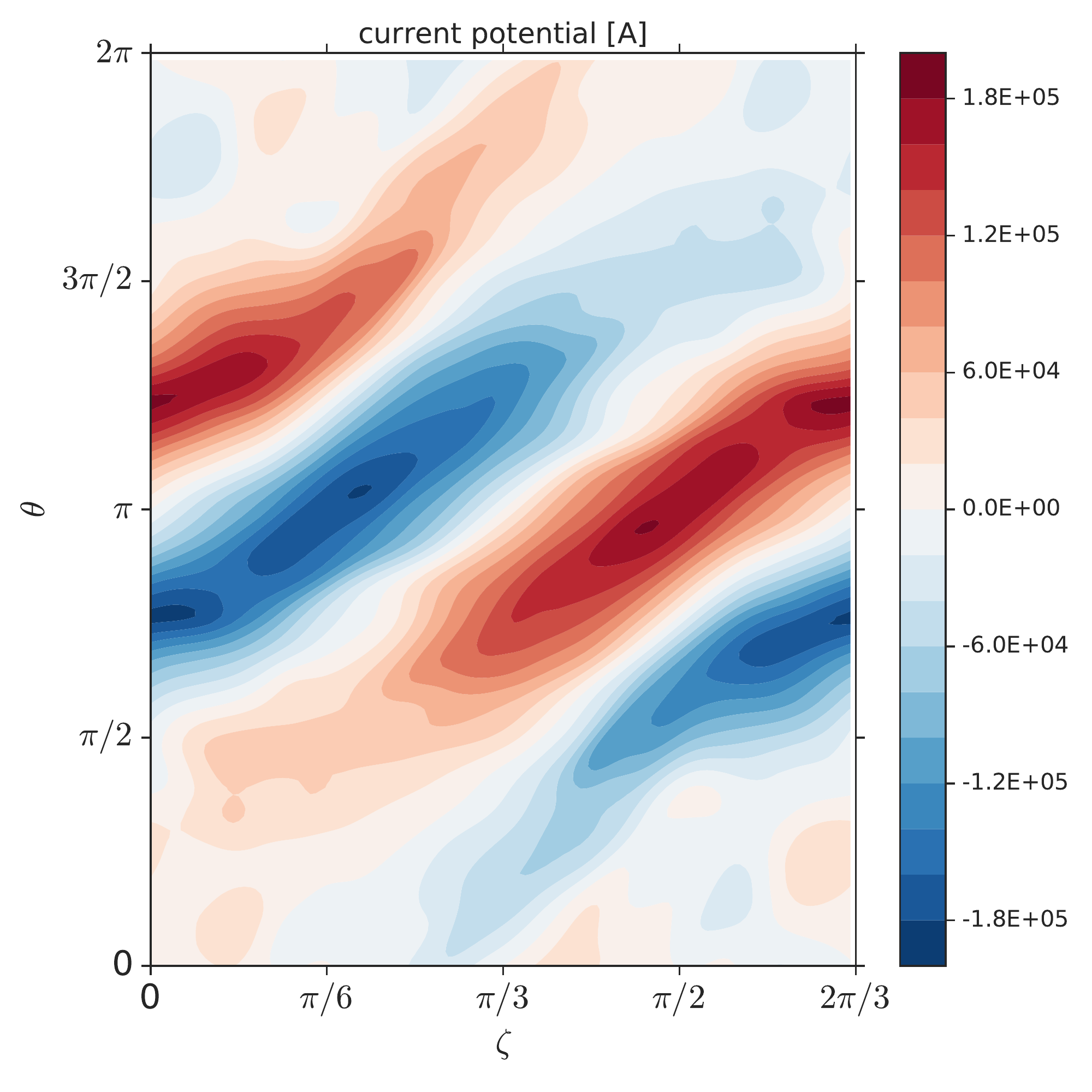}
    \caption{Contour plot of the current potential solved on the vacuum vessel for the half-Tesla NCSX C09R00 configuration. The regularization parameter $\lambda$ is $1.46\tento{-17}$ selected at the corner of the "L" curve. }
    \label{fig:ncsx_cp}
\end{minipage}\hfill
\begin{minipage}[t]{.49\linewidth}
    \centering
    \includegraphics[width=\linewidth]{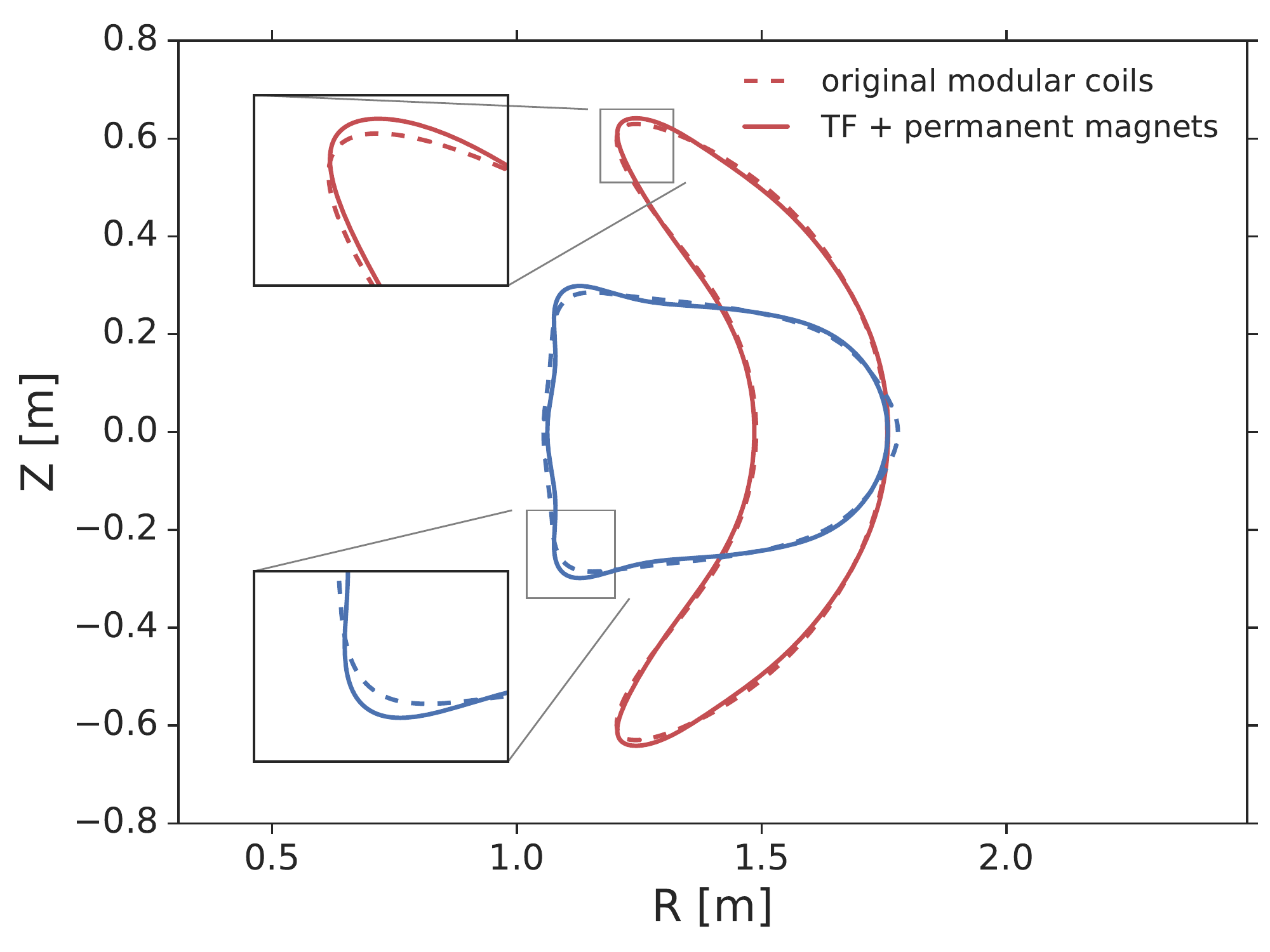}
    \caption{Plasma boundaries of free-boundary VMEC calculations with the magnetic field produced by the original modular coils (dash lines) and by the TF coils together with permanent magnets (solid lines), respectively. The two symmetry cross-sections, $\z = 0$ (bean-shaped, red lines) and  $\z = \pi/3$ (bullet-shaped, blue lines), are shown. }
    \label{fig:ncsx_fb}
\end{minipage}
\end{figure}

The actual NCSX vacuum vessel was used as the winding surface and the toroidal field was generated by 18 TF coils \change{which}{that} have been built. 
Figure \ref{fig:ncsx_cp} shows the required current potential on the winding surface to cancel off the residual magnetic field produced by the TF coils and non-zero plasma currents.
This is calculated by using REGCOIL regularized with the current density and the residual field error $\chi^2_B$ is $1.49\tento{-5} \mathrm{T^2m^2}$ at the corner of the ``L" curve.
The maximum value of the current potential is about $1.86\tento{5}$ A, which is equivalent to \change{the surface magnetic moment}{the magnetic moment per unit area} of 17-cm-thick Nd-Fe-B magnets.
The minimum distance from the vacuum vessel to the plasma boundary is even smaller than the required thickness of magnets at some locations.
Hence, the surface moment approximation is not sufficient and it is necessary to carry out a finite-thickness calculation following algorithm \ref{alg1}.
%

\begin{figure}
    \centering
    \includegraphics[width=\linewidth]{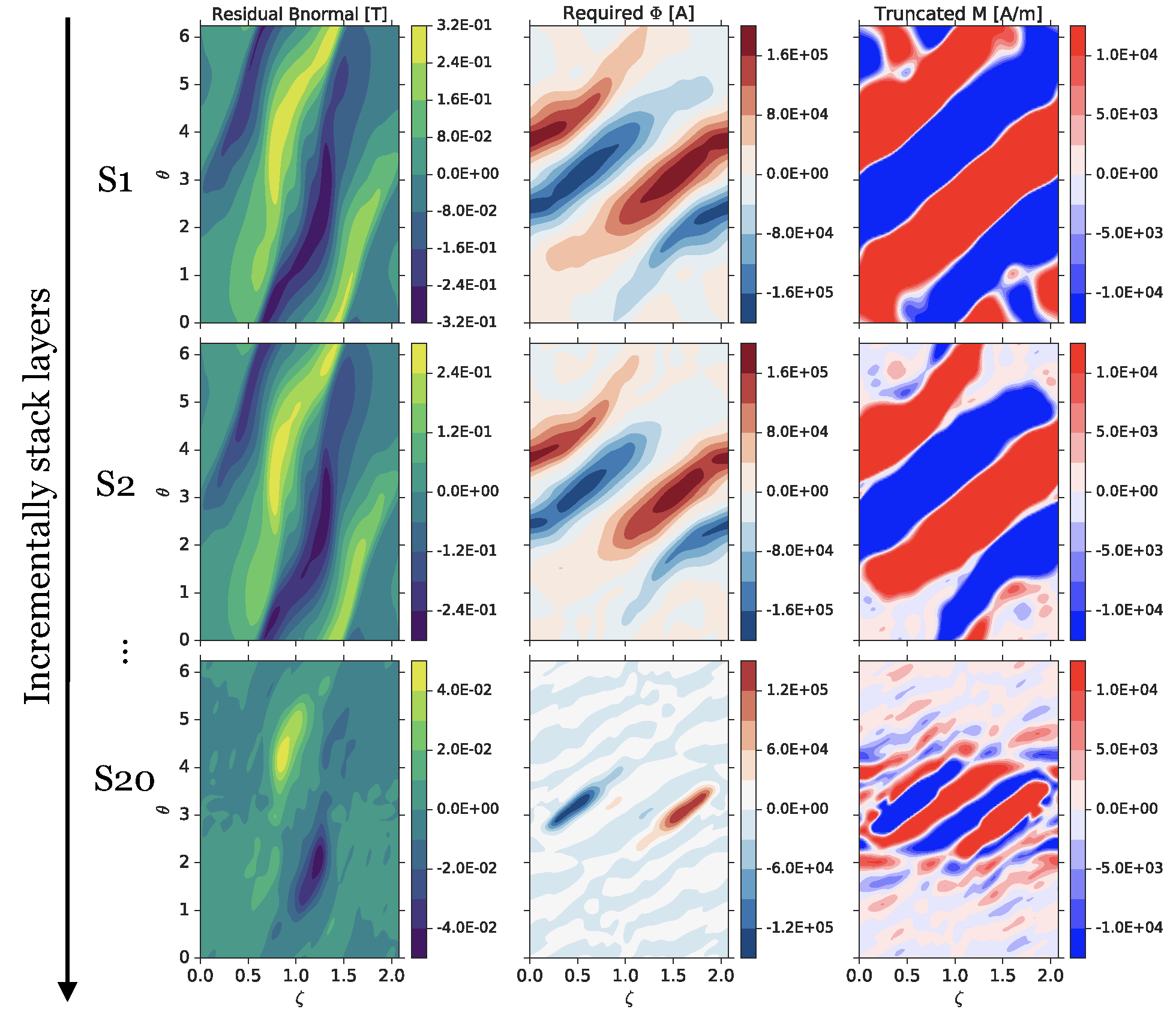}
    \caption{\change{}{Demonstration of the multi-layer implementation for NCSX. The innermost layer, $S_1$, is the vacuum vessel and nested surfaces with a distance of 1 cm, $S_2 \cdots S_{20}$ are used. The residual normal field to be canceled off (left), the solved surface current potential (middle), and the truncated magnetization (right) are shown for $S_1$ (the first row), $S_2$ (the second row) and $S_{20}$ (the third row).}}
    \label{fig:demo}
\end{figure}

\begin{figure}
    \centering
    \includegraphics[width=0.95\textwidth]{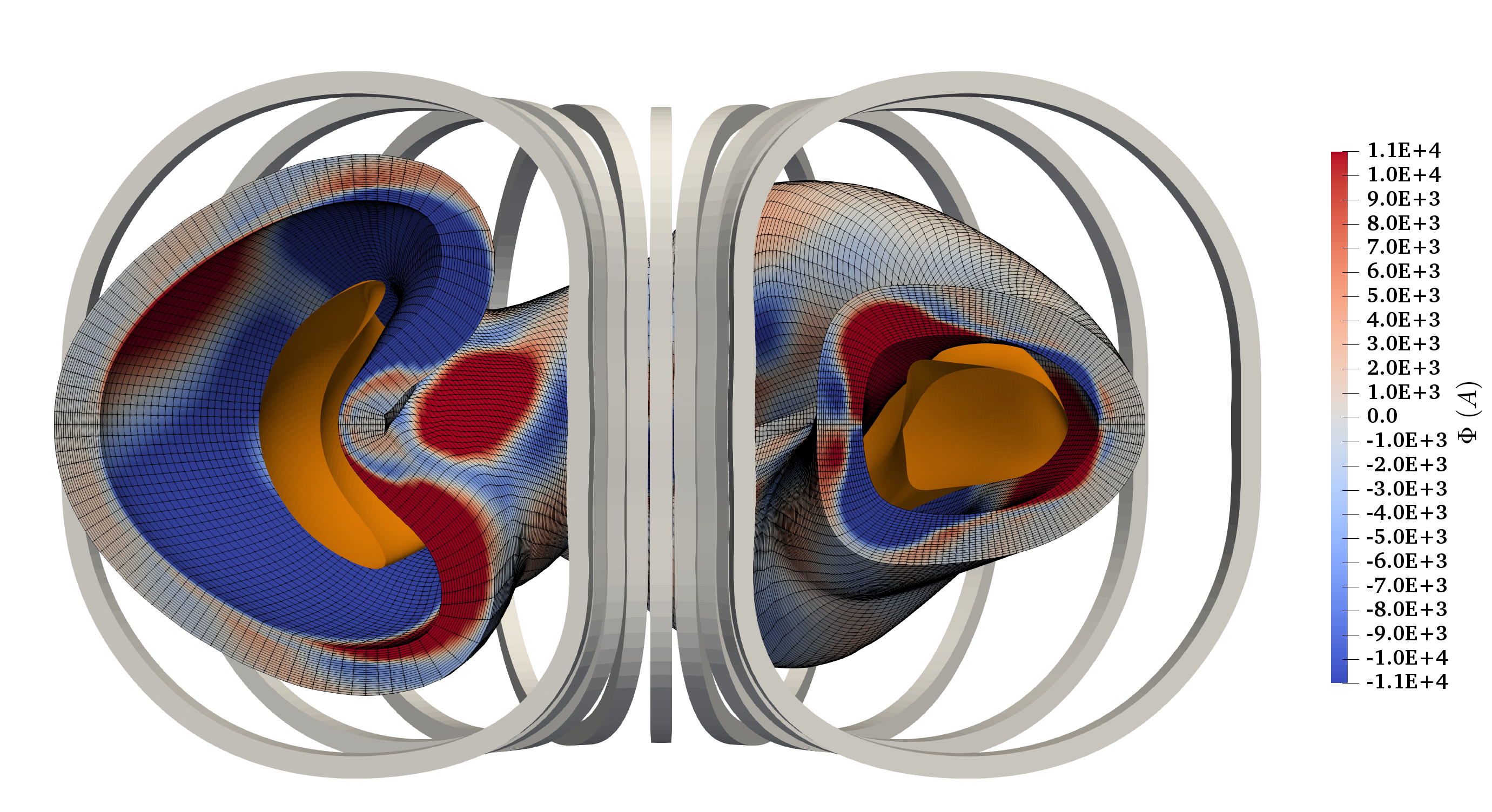}
    \caption{The half-Tesla NCSX configuration with planar TF coils and the permanent magnets. Only half of the torus is shown. The C09R00 plasma boundary \change{are}{is}  shown in orange. Colors in the computation domain indicate the magnitude of the magnetization of the discretized permanent magnet elements and positive values mean pointing inwards. The grey regions have negligible magnetization and can be omitted.}
    \label{fig:ncsx_final}
\end{figure}

The vacuum vessel was used as the starting surface ($S_1$) and other surfaces were generated by uniformly expanding around the vacuum vessel surface.
The distance between two adjacent surfaces was scanned over 1 mm, 2
mm, 4 mm, 6 mm, 8 mm and 1 cm.
Scan results show that a radial resolution of 1 cm is adequate.
The last surface ($S_{20}$) was chosen to be 20 cm away from the vacuum vessel because the concave region at the bean-shaped cross-section began to overlap when the surface was expanded further than 20 cm along the surface normal vector.
\change{}{The residual normal field to be canceled off, the solved surface current potential, and the truncated magnetization at each iteration are shown in \Fig{demo}.
The required current potential becomes less when more layers are stacked.
On the last layer, $S_{20}$, the required current potential is centralized at the inboard side and it is actually not converged.}
The new design using only planar TF coils and finite-thickness permanent magnets is illustrated in \Fig{ncsx_final}.
The resolution of the magnetic dipoles is $N_d^r \times N_d^{\t} \times N_d^{\z} = 20 \times 128 \times 128$ and all the dipoles are orientated perpendicularly to the surface that they are lying on.
\change{Magnetic moment of the dipoles are calculated as the product of
surface magnetization and the thickness}{The magnetic moment of each dipole is calculated as the product of current potential and the element area} by the following \Eqn{dipole}.
\change{The maximum current potential is limited to $1.1\tento{4}$ A, which is the surface magnetization of 1-cm-thick Nd-Fe-B magnet.}
{The surface magnetization is truncated with a maximum value of $1.1\tento{4}$ A/M to approximate 1-cm-thick Nd-Fe-B magnets.}

With such 20-cm-thick permanent magnets, the residual field error $\chi^2_B$ is $9.02\tento{-5} \mathrm{T^2m^2}$ ( average $|\vect{B} \cdot \vect{n}|/{|\vect{B}|}$ $5.23 \tento{-3}$). 
It is slightly higher than the one calculated by \change{}{a} single layer \change{}{of} current potential, since we stopped at 20 cm and the field error is not converged.
However, the free-boundary VMEC \cite{VMEC_fb} calculations, shown in \Fig{ncsx_fb}, are close to the ones from the original modular coils.
The average relative difference in rotational transform profiles is 1.79\%.
From the perspective of the magnetic field, \change{this}{the accuracy} is acceptable.

The total magnetic moment used here is equal to an Nd-Fe-B magnet of $3.6 \ \mathrm{m^3}$. 
The cost \change{for}{of}  purchasing such amount of magnet materials would be at the order of million U.S. dollars, which is remarkably attractive.
\change{Of course, other costs, like assembling the magnets, should also be assessed when designing a real permanent magnet stellarator in the future.}
{Other costs, like the supporting structure and assembling cost, might be more expensive than the material.
Nevertheless, thorough cost comparisons should be performed when a real permanent magnet stellarator is going to be built.
}

\section{Discussion and Conclusions} \label{discussion}
In this paper, we introduce a multi-layer method to design perpendicular permanent magnets for stellarators.
\change{The surface current is proven to be equivalent to a surface magnetization.}
{The surface current potential is proven to be equivalent to the dipole moment per unit area.}
With the layer-by-layer incremental calculation, we can design permanent magnets with finite thickness to ensure the maximum magnetization is achievable with present materials.
Numerical validation on the rotating elliptical stellarator demonstrates that the permanent magnets (together with a toroidal field) can provide a precise magnetic field for confining plasmas.
The multi-layer method is then applied to design permanent magnets for the QA stellarator NCSX.
The new design using only planar TF coils and permanent magnets shows that by using permanent magnets it is possible to build optimized stellarators with extremely simple coils.
Besides the potential cost \change{reduce}{reduction}, using permanent magnets could likely provide huge access on the outboard side.
As shown in \Fig{ncsx_final}, for NCSX, most of \change{}{the} permanent magnets are locating at the inboard side, while only one or two layers are needed on the outboard side.

The method uses the surface current potential which could be linearly solved by using NESCOIL or REGCOIL.
Therefore, it is fast and robust.
The multi-layer solution could be a relatively good initial guess for nonlinear optimization codes.
This is not trivial as local optimization methods are more commonly used in stellarator optimizations than global minimization algorithms for the consideration of speed.
The nonlinear optimization code to design permanent magnets is expected to be easily trapped into local minima due to the extremely high dimensionality. 
\change{We can also use the results as a proxy in stellarator optimization codes, like STELLOPT and ROSE, to help find new configurations that requires simple coils and light permanent magnets.}{}
Since the magnetic field produced by the surface magnetization is essentially identical to the one generated by the surface current, this approach can be applied to equilibrium reconstruction without designing modular coils \cite{Mikhailov2017}.

The multi-layer method is not an overall optimization, as the magnetization on inner layers will not be affected by outer ones.
This is why reversed magnetization appears at the inboard side of the bullet-shaped cross-section in \Fig{ncsx_final}, which could be eliminated using overall optimization methods.
Only the magnets that are perpendicular to the winding surface are used in this paper.
By doing this, it might be easier for future engineering designs.
We can replace nearby dipoles that have the same orientation with larger pieces of magnets.
\change{Meanwhile, relaxing the orientation would give us more freedom to find different solutions, especially it can reduce the total volume of required magnets.
Exploration on using non-perpendicular magnets will be addressed in future work.}{}
To avoid ports and other engineering features, it is possible to target stay-away zones in the optimization by adding a regularization with nonuniform spatial weights on the current potential.
\change{In this paper, we haven't considered any engineering constraints, like mounting/supporting structures, magnetic forces for each magnet, magnetic hysteresis, etc.
These calculations might require advanced engineering analysis tools, which are beyond the scope of this paper.}
{}

\change{}{
There are also challenges for using permanent magnets in stellarators.
For example, the limitation on the field strength is constrained by the remanent field of the material, which is usually less than 2 T.
This is not enough for future fusion reactors.
\changenew{}{It will be even more challenging considering the fact that the magnets have to be placed outside the vacuum vessel which will be at least 1 meter away from the plasma.}
\changenew{}{The feasibility of permanent magnets for reactor-size fusion experiments with similar neutron flux needs further investigations.}
However, we don't have to use permanent magnets to produce all the rotational transform, although in the above configurations external coils only produce the toroidal field.
For possible future experiments with permanent magnets, we would envision that permanent magnets will be used to generate a certain percentage of rotational transform until the coil complexity is tolerable.
The other fact that the half-Tesla NCSX requires 20-cm-thick magnets needs further discussions.
First of all, as we stated, relying on permanent magnets to generate the entire rotational transform, which is the case for NCSX, is not the best solution.
Second, only perpendicular magnets have been considered.
Relaxing the orientation would give us more freedom to find different solutions, especially it can reduce the total volume of required magnets.
Exploration on using non-perpendicular magnets will be addressed in future work.
Third, the NCSX configuration was optimized to use modular coils, not permanent magnets.
Actually, the difference of required magnets between the rotating ellipse and NCSX is distinct.
The rotating ellipse has a higher field ($B_{axis}=0.67$ T) but requires much fewer magnets ($\sim$ 3 cm thick).
The linear method proposed in this paper can be used as a proxy in stellarator optimization codes, like STELLOPT \cite{STELLOPT} and ROSE \cite{ROSE}, to help find new configurations that require simple coils and light permanent magnets.
Some important engineering considerations, like magnetic forces, supporting structures, assembly tolerance, nonlinear magnetic permeability, etc., should be evaluated at the stage of engineering design when a real experiment is going to be built.}


\section*{Acknowledgments}
CZ gratefully appreciates fruitful discussions with Peifeng Fan, Matt Landreman, Elizabeth Paul, Tony Qian, Per Helander and Steven Cowley.
This work was supported by the U.S. Department of Energy under Contract No. DE-AC02-09CH11466 through the Princeton Plasma Physics Laboratory.

\begin{appendices}
\numberwithin{equation}{section}

\section{Details to obtain \Eqn{AK}} \label{derivation}
Substituting \Eqn{current_density} into \Eqn{bs_surface}, we have
\begin{align}
    \ds \vect{A}_K & = \frac{\mu_0}{4\pi} \iint_{\partial C} \frac{\vect{n} \times \grad{\Phi}}{r} \dd{S} \ \nonumber \\
                   & = \frac{\mu_0}{4\pi} \iint_{\partial C} \qty [ \Phi \curl{\qty(\frac{\vect{n}}{r})} - \curl{\qty(\frac{\Phi \vect{n}}{r}}) ] \dd{S} \  \nonumber \\
                   & = \frac{\mu_0}{4\pi} \iint_{\partial C} \qty [ \Phi \grad{\frac{1}{r}} \times \vect{n} + \frac{\Phi}{r} \curl{\vect{n}}  - \curl{\qty(\frac{\Phi \vect{n}}{r}}) ] \dd{S} \ \nonumber \\
                   & = \frac{\mu_0}{4\pi} \iint_{\partial C}  \qty [ - \Phi \frac{\vect{r}}{r^3} \times \vect{n} - \grad{\qty(\frac{\Phi}{r})} \times \vect{n} ] \dd{S} \ \nonumber \\
                   & = \frac{\mu_0}{4\pi} \iint_{\partial C} \frac{\Phi \vect{n} \times \vect{r}}{r^3} \dd{S} + \frac{\mu_0}{4\pi} \iint_{\partial C} \vect{n} \times \grad{\qty(\frac{\Phi}{r})} \dd{S} \nonumber \\
                   & = \frac{\mu_0}{4\pi} \iint_{\partial C} \frac{\Phi \vect{n} \times \vect{r}}{r^3} \dd{S} + \frac{\mu_0}{4\pi} \iiint_{C} \qty[\curl{\grad{\qty(\frac{\Phi}{r})}}] \dd{V} \nonumber \\
                   & = \frac{\mu_0}{4\pi} \iint_{\partial C} \frac{\Phi \vect{n} \times \vect{r}}{r^3} \dd{S} \ .
\end{align}



\section{Regularization over the current potential} \label{phi_regularize}
New regularization term over the integral of current potential squared, 
\begin{align}
    \chi^2_{\Phi} & = \int_{S} \dd{S} \Phi^2  \ ,
\end{align}
is implemented in REGCOIL. The total current potential is expressed as,
\begin{equation}
    \Phi = \int_{S} \dd{S} \qty(\sum_j \Phi_j p_j + \frac{G}{2\pi} \zeta + \frac{I}{2\pi} \theta)^2 \ ,
\end{equation}
where $p_j$ is the basis function,
\begin{equation}
    p_j = 
    \begin{pmatrix}
    \sin \\
    \cos \\
    \end{pmatrix}_j
    (m_j \t - n_j \z) \ .
\end{equation}
To solve $\partial \chi^2_{\Phi} / \partial \Phi_j = 0$, we could obtain similar linear equation as in REGCOIL original paper \cite{REGCOIL}, and the matrices are
\begin{equation}
    A^\Phi_{j,k} = \Delta \t \Delta \z \sum_{i_t} \sum_{i_\z} \qty( {N} p_j p_k)
\end{equation}
\begin{equation}
    b^\Phi_j = - \Delta \t \Delta \z \sum_{i_t} \sum_{i_\z} \qty[ {N} p_j \qty( \frac{G}{2\pi} \zeta + \frac{I}{2\pi} \theta)]
\end{equation}

\begin{figure}
\centering
\begin{minipage}[t]{.49\linewidth}
    \centering
    \includegraphics[width=\textwidth]{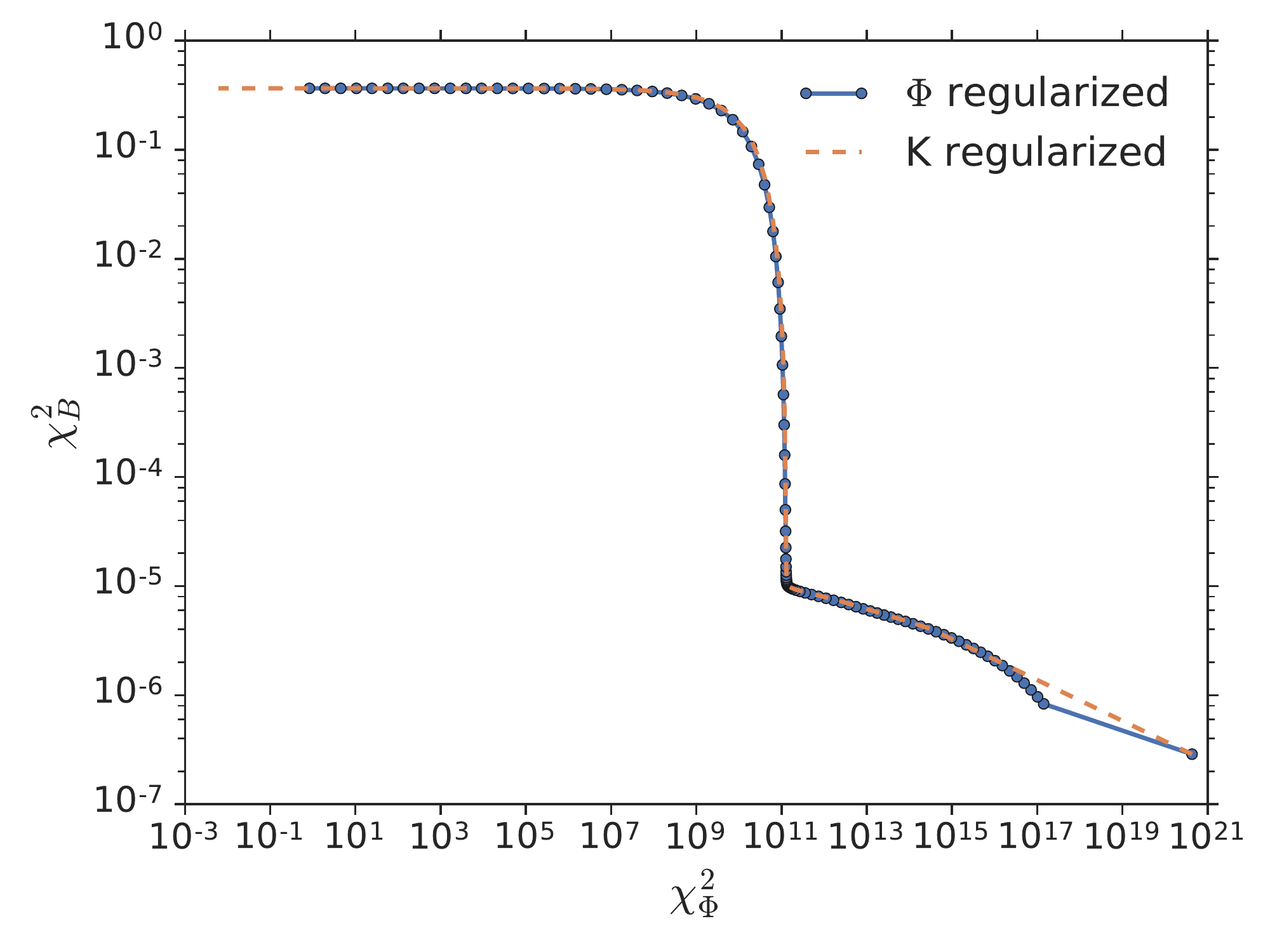}
    \caption{\change{}{Regularization curves of $\chi^2_B$ with respect to $\chi^2_{\Phi}$ when scanning the regularization parameter with the current density regularization or the current potential regularization.} }
    \label{fig:phi_reg}
\end{minipage}\hfill
\begin{minipage}[t]{.49\linewidth}
    \centering
    \includegraphics[width=\linewidth]{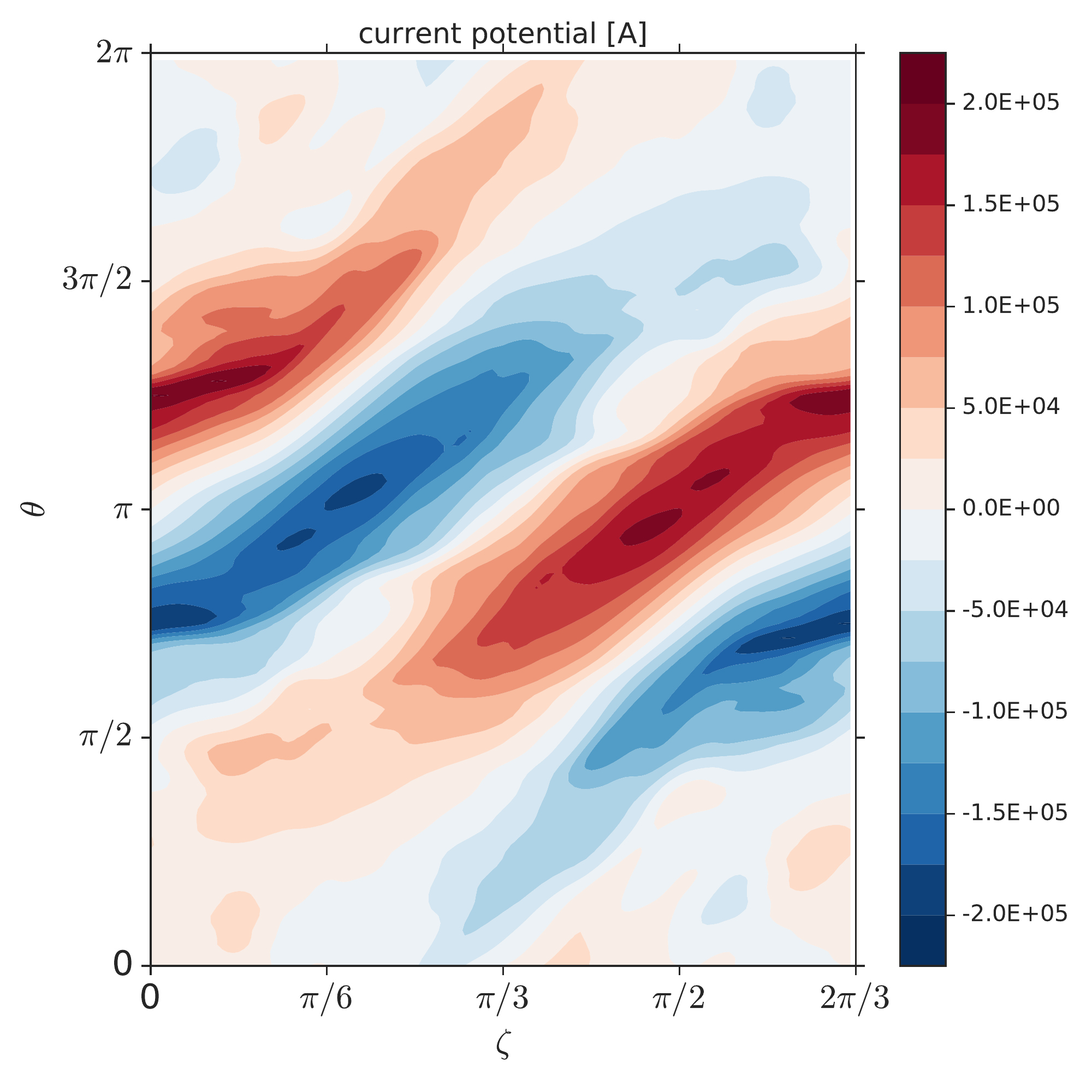}
    \caption{\change{}{Contour plot of the current potential solved by using the current potential regularization with $\chi^2_B = $. All the other settings are identical to the one shown in \Fig{ncsx_cp}.}}
    \label{fig:ncsx_cp_phi}
\end{minipage}
\end{figure}

The original regularization term, current density, is penalizing the solution with the gradient of $\Phi$.
It also has a influence on the magnitude of $\Phi$ and likely provides a smoother solution than using the current potential regularization itself.
\change{}{Figure \ref{fig:phi_reg} shows the $\chi^2_{\Phi}$ -$\chi^2_B$ curves when we scanned the regularization parameter $\lambda$ using the regularization over the current density or the current potential.
The two curves are almost identical and prove that the current density regularization can also regularize the current potential.}
When we use the new regularization term over the current potential, we tend to get solutions with more localized peaks.
\change{}{As shown in \Fig{ncsx_cp_phi}, we selected one solution from the scanned results using the current potential regularization.
It has a close value of $\chi^2_B$ ($1.50\tento{-5} \mathrm{T^2m^2}$) compared to the one in \Fig{ncsx_cp} ($\chi^2_B = 1.49\tento{-5} \mathrm{T^2m^2}$), but the maximum current potential is significantly higher and more localized.}





\end{appendices}

\section*{References}
\bibliography{magnets}

\end{document}